\documentclass[bibauthoryear]{aa}

\usepackage{amssymb}
\usepackage{soul}

\newcommand{\der}{\mathrm{d}}

\usepackage{graphicx}
\usepackage[varg]{txfonts}
\usepackage{amsmath}
\usepackage{natbib}
\usepackage{xcolor}
\usepackage[colorlinks=true,citecolor=blue,linkcolor=blue,urlcolor=blue,anchorcolor=blue]{hyperref}

\begin{document}

	\title{A quantitative explanation of the cyclotron-line variation in accreting magnetic neutron stars of super-critical luminosity}
	
	\author{Nick Loudas
		\inst{1,2,3}
		\and
		Nikolaos D. Kylafis\inst{1,2}
		\and
		Joachim Tr\"{u}mper\inst{4,5}
	}
	
	\institute{
    University of Crete, Department of Physics \& Institute of
		Theoretical \& Computational Physics, 70013 Herakleio, Greece
              \\
             \email{kylafis@physics.uoc.gr}
		\and
		Institute of Astrophysics,
		Foundation for Research and Technology-Hellas, 71110 Heraklion, Crete, Greece
            \and
            Department of Astrophysical Sciences, Peyton Hall, Princeton University, Princeton, NJ 08544, USA \\
		\email{loudas@princeton.edu}
		\and 
		Max-Planck-Institut f\"{u}r extraterrestrische Physik, 
		Postfach 1312, 85741 Garching, Germany
        \and
        University Observatory, Faculty of Physics, Ludwig-Maximilians Universit\"{a}t, Scheinerstr. 1, 81679 Munich, Germany\\
	}
	
	\date{Received / accepted }
	
\abstract
{
 Magnetic neutron stars (NSs) often exhibit a cyclotron resonant scattering feature (CRSF) in their X-ray spectra. Cyclotron lines are believed to be generated in the radiative shock in the accretion column. High-luminosity NSs show a smooth anti-correlation between the cyclotron-line centroid ($E_{CRSF}$) and X-ray luminosity ($L_X$). 
 }
{
It has been pointed out that the observed $E_{CRSF}-L_X$ smooth anti-correlation in high-luminosity NSs is in tension with the theoretically predicted one, if the radiative shock is the site of cyclotron-line formation. The shock height increases approximately linearly with luminosity, while the dipolar magnetic field drops as the cubic power of distance, thereby implying that the cyclotron-line energy ought to decrease significantly when the luminosity increases by, say, an order of magnitude, which is contrary to observations. Since there is no other candidate site for the cyclotron-line formation, we re-examine the predicted rate of change of the cyclotron-line energy with luminosity at the radiative shock, taking a closer look at the Physics involved.
 }
	{We developed a purely analytical model describing the overall dependence of the observed cyclotron energy centroid on the shock front's height, including both the relativistic boosting and the gravitational redshift effects in our considerations. The relativistic boosting effect is due to the mildly relativistic motion of the accreting plasma upstream with respect to the shock's reference frame. 
 }
	{
 We find that the cyclotron-line energy varies with a) the shock height due to the dipolar magnetic field, b) the Doppler boosting between the shock and bulk-motion frames, and c) the gravitational redshift. We show that the relativistic effects noticeably weaken the predicted $E_{CRSF}-L_X$ anti-correlation.
 We use our model to fit the data of the X-ray source V0332+53 that exhibits a weak negative correlation and demonstrate that the model fits the data impressively well, thereby alleviating the tension between observations and theory. 
 }
{
The reported weak anti-correlation between cyclotron-line centroid and X-ray luminosity in the supercritical accretion regime may be explained by the combination of the variation of the magnetic-field strength along the accretion column, the effect of Doppler boosting, and the gravitational redshift. As a result of these effects, the {{\it actual}} magnetic field on the surface of the neutron star may be a factor of $\sim 2$ larger than the naively inferred value from the observed CRSF.
 }
	

	\authorrunning{N. Loudas et al.}
	\titlerunning{Cyclotron line variation in the super-critical regime} 
	
	\maketitle
	
	
\section{Introduction} \label{sec1}

	Accretion-powered X-ray pulsars (XRPs) are highly magnetized neutron stars (NSs) in binary systems. The strong dipole-like magnetic field ($B\sim \text{a few} \times 10^{12}$ G) channels the accreting plasma onto the magnetic poles, where most of the accretional energy is released in X-rays (\citealt{Basko1975,Basko1976}). Their X-ray spectra are generated through bulk- and thermal-Comptonization and are generally characterized by a power law followed by a hard-photon tail and an exponential cut-off \citep{Lyubarskii1982}. For a recent review, see \cite{Mushtukov2022}.

    A substantial fraction of them exhibit absorption-line-like features in their high-energy X-ray spectra, called cyclotron lines or cyclotron resonant scattering features (CRSFs). The underlying physical process that is responsible for the formation of such features is a purely quantum-mechanical phenomenon called magnetic resonant scattering. The motion of a charged particle (e.g., electron) under the influence of a strong magnetic field is quantized perpendicular to its direction (\citealt{Landau1930}; see also \citealt{Landau1965}), leading to discrete energy states that are commonly known as Landau levels. Hence, absorption/emission lines are expected to appear in the spectra of highly magnetized NSs. 
    
    Detecting a CRSF in the spectrum of a NS provides a clear way of measuring its magnetic-field strength \citep{Gnedin1974} and provides an excellent diagnostic tool to explore the physics of accretion onto the star \citep{Staubert2019}.
    The difference in energy between two successive levels is directly proportional to the magnetic-field's strength in the line-forming region. The energy of the fundamental line (i.e., the transition from the ground state to the first excited one) is expected to be at about the classical cyclotron energy
    
    \begin{equation}
    E_c = \dfrac{\hbar e B}{m_e c} \approx 11.6 \left(\dfrac{B}{10^{12}~\mathrm{G}}\right) ~\mathrm{keV}, \label{1.1}
    \end{equation}
    where $\hbar$ is the reduced Planck's constant, $m_e$ is the electron's rest mass, $c$ is the speed of light, $e$ is the magnitude of the electron's charge, and $B$ is the magnetic-field strength in the line-forming region.
    
    The first-ever cyclotron line was detected in the X-ray spectrum of the accreting binary pulsar Hercules X-1 by \cite{Truemper1977} \citep[see also][]{Truemper1978}. Since then, more than 35 electron cyclotron lines have been identified in accreting magnetic NS X-ray spectra, sometimes with their harmonics and even more rarely with anharmonics (see e.g., \citealt{Fuerst2018,Sharma2022,Yang2023}), covering an energy range from $\sim 10~\mathrm{keV}$ to $\sim 100~\mathrm{keV}$. 
    
    In a considerable fraction of these sources, the line centroid appears to vary noticeably with the observed X-ray luminosity $L_X$. For a thorough review, see \cite{Staubert2019}. High-luminosity sources ($L_X \gtrsim 10^{37} ~\mathrm{erg ~ s^{-1}}$) exhibit a negative correlation between the observed line energy $E_{\rm CRSF}$ and the emergent X-ray luminosity $L_X$ (e.g., V0332+53: \citealt{Tsygankov2006}), while low-luminosity sources ($L_X \lesssim 10^{37} ~\mathrm{erg~ s^{-1}}$) show a positive one (e.g., Her X-1: \citealt{Staubert2007}; GX 304-1: \citealt{Klochkov2012}; Vela X-1:  \citealt{Fuerst2014}). Whether the correlation is positive or negative seems to be dictated by a critical luminosity value ($L_{cr} \sim 10^{37}\,\mathrm{erg ~ s^{-1}}$) above which the correlation becomes negative, i.e., an anti-correlation (see, e.g., \citealt{Shui2024}). This anti-correlation was predicted by \cite{Basko1976} and has already been observed in two XRPs (1A 0535+262: \citealt{Kong2021,Shui2024} and V 0332+53: \citealt{Doroshenko2017,Vybornov2018}), but see also GRO J1008-57 \citep{Chen2021}.

    In sources of relatively low or sub-critical luminosity, the accreting plasma is believed to almost reach the NS surface without braking, and the final braking, as well as the formation of the CRSF occurs in the NS atmosphere.
    In the literature, two physical models have been proposed to explain the plasma braking in the atmosphere and the positive $E_{\rm CRSF} - L_{X}$ correlation reported for these sources. The first one is associated with the emergence of a collisionless shock (CS) above the polar cap, which dissipates the ions' kinetic energy (e.g., \citealt{Bisnovatyi1970,Shapiro1975}). The CS's height decreases as the accretion rate (luminosity) increases (see e.g., \citealt{Bykov2004}), and the magnetic-field strength (cyclotron energy) also increases with the CS height's drop. Thus, a positive $E_{\rm CRSF} - L_X$ correlation is expected (for details see \citealt{Becker2012}). For an implementation, see, e.g., \cite{Staubert2007, Vybornov2017,Rothschild2017}. 
    
    The second model, proposed by \cite{Musthukov2015b}, does not invoke a CS, but suggests that the braking of the plasma occurs via multiple scatterings of the accreting electrons with the photons from the hot-spot, at a height comparable to the size of the hot spot.  The picture is as follows.
    Up-ward going X-ray photons undergo resonant scattering with the accreting electrons, resulting in the emergence of a CRSF that is red-shifted with respect to the local cyclotron line energy $E_c$, because of the Doppler shift between the lab fame (hot spot) and the moving frame (accreting electrons). The radiation pressure reduces the velocity of the accretion-flow, thus affecting the shift of the CRSF. The higher the luminosity (radiation pressure), the smaller the accretion-flow velocity (redshift), and thus the larger the energy of the observed CRSF centroid.
    The model has been successful in describing the data of the XRP GX 304-1. \cite{Markozov2023} advanced the model and obtained detailed velocity profiles and emergent spectra, accounting for resonant scattering.
    Yet, no quantitative calculation has been performed so far for the variation of cyclotron lines in these models.

    In contrast, the accretion picture is different for sources of high luminosity, that often exhibit a negative $E_{\rm CRSF} - L_X$ correlation. In this luminosity regime, the role of the photons is different. The outgoing radiation can create and sustain a radiation-dominated shock (RS) in the accretion column, which is responsible for the braking of the in-falling plasma \citep{Basko1975,Basko1976,Becker2007,Becker2012,Zhang2022}. The formation of the CRSF is naturally expected to occur at the RS. The height of the RS scales approximately linearly with the X-ray luminosity, while the magnetic field strength drops with the increase of the shock's height. Thus, it predicts an anti-correlation, which is qualitatively in line with the observations of super-critical XRPs. 
    
    \cite{Poutanen2013} pointed out that, due to the steep $1/r^3$-dependence of the dipole magnetic field with distance $r$ (see Eq. \ref{2.1} below), the rate of change of $E_{\rm CRSF}$ with $L_X$ should be much larger than the observed one.
    Instead, they proposed that the observed CRSF is not generated in the RS, but rather by
    reflection of radiation (produced in the RS) in the area surrounding the polar cap. 
    The negative correlation can, in principle, be obtained by the reprocessing of photons on the NS surface. As the luminosity increases, the height of the RS  also increases, and thus a larger fraction of photons are scattered at lower magnetic latitudes, where the magnetic field is lower (the magnetic field at the equator is half that of the pole). \cite{Kylafis2021} explored this scenario by means of Monte Carlo (MC) calculations and demonstrated that this mechanism could neither produce prominent CRSFs in the emergent X-ray spectra nor explain the $E_{\rm CRSF} - L_X$ anti-correlation reported in the  V0332+53 source (\citealt{Tsygankov2010}).  Adding the spectra from all illuminated magnetic latitudes washes out the expected correlation.

    In \cite{Loudas2023}, we revisited the model of \cite{Basko1976} and calculated the X-ray spectrum produced in the RS of super-critical XRPs by means of MC simulations, implementing a fully relativistic scheme and employing accurate magnetic resonant scattering cross-sections to allow for a CRSF to arise. We demonstrated that bulk- and thermal-Comptonization, as well as resonant scattering of photons by electrons in a RS,  result naturally in a power-law X-ray continuum with a high-energy cutoff, in line with previous theoretical works \citep{Lyubarskii1982}.  We showed that a RS is efficient in producing a prominent CRSF in the emergent X-ray spectrum. Moreover, we highlighted the implications of the Doppler effect on the centroid of the emergent CRSF and found that the cyclotron-line energy centroid is shifted by $\sim(20-30)\%$ to lower energies compared to the classical cyclotron energy $E_c$, due to the Doppler boosting between the stationary, shock reference frame, and the bulk-motion one. 
    
    As a natural extension of our previous work \citep{Loudas2023}, in this study we re-examine the predicted rate of change of $E_{\rm CRSF}$ with $L_X$ in the RS model, and quantify the importance of the Doppler effect and the gravitational redshift on the anti-correlation.  Thus, we reconsider
    the argument of \cite{Poutanen2013} about the  
    predicted $E_{\rm CRSF} - L_X$ correlation in the RS model.
	
    The layout of this paper is as follows. In Sect. \ref{sec2} we introduce our model, in Sect. \ref{sec3} we demonstrate our results, providing a quantitative comparison with observations of the source $V0332+53$, and in Sect. \ref{sec4} we comment on the results and draw our conclusions.

\section{The model} 
\label{sec2}

\subsection{Standard picture}
\label{sec2.1}

\subsubsection{Accretion geometry}
\label{2.1.1}

We envision the accretion flow and the column geometry in accreting magnetic NSs of super-critical luminosity as follows. The strong dipole magnetic field channels the accretion disk's plasma to the magnetic poles of the NS, forming an accretion column above each pole. We assume the accretion column to be a cylinder of cross-sectional area $\pi a_0^2$ (though, accretion physics dictates that the accretion column is not cylindrical, but bow-like around the magnetic pole; see, e.g., \citealt{Basko1976, Romanova2004}) and a radiative shock at height $H$ above the magnetic pole. Above the shock, the matter is in free fall (pre-shock region), while below the shock (post-shock region) it has negligible flow velocity
(one seventh the free-fall one), density seven times the pre-shock one (appropriate for radiative strong shocks; \citealt{Blandford1981}), and the electrons' motion is mainly thermal. 

The free-falling plasma is decelerated by the radiation field while passing through the shock. The shock is treated here as a mathematical discontinuity, though, in reality, radiation-mediated shocks are continuous having a finite size on the order of a few photon mean free path (e.g., Eq. 15 in \citealt{Loudas2023}). In \cite{Loudas2023}, we have proven that the thickness of the radiative shock in the accretion column is way smaller than its altitude from the NS surface.  The accreting plasma's kinetic energy is supplied to photons via multiple electrons-photon scatterings. Ions carry most of the accretional energy and electrons act as mediators in transferring this energy to photons. In this way, the accreting plasma loses most of its kinetic energy and eventually sinks slowly in the post-shock region. A detailed description of the deceleration process is given in Sect. 2.1.2 of \cite{Loudas2023}.

\subsubsection{Standard cyclotron-line formation}
\label{2.1.2}
According to the proposed model of \cite{Basko1976}, the formation of cyclotron lines in NSs of super-critical luminosity occurs at the RS in the accretion column. The cyclotron-line energy centroid depends on the local magnetic-field strength in the line-forming region (i.e., the shock). 

We assume a dipole magnetic field, whose strength at a distance $r$ from the NS center is
\begin{equation}
    B(r) = B_* \left(\dfrac{R_*}{r}\right)^3 ,
    \label{2.1}
\end{equation}
where $R_*$ is the NS radius and $B_*$ is the magnetic field strength on its surface.  The cyclotron-line energy at the shock's altitude $H$ is then
\begin{equation}
    E_c (H) = E_{c,*} \left(\dfrac{1}{1 + H/R_*}\right)^3 ,
    \label{2.2}
\end{equation}
where $E_{c,*}$ denotes the cyclotron-line centroid energy on the NS surface.   

The shock's altitude depends on the mass accretion rate and thus on the emergent luminosity. \cite{Basko1976} demonstrated that $H$ scales approximately linearly with $L_X$ (see also, \citealt{Becker2012}). Hence, we can write down the following scaling relation for $L_X > L_{cr}$
\begin{equation}
    L_X = L_{cr} \left(\dfrac{H}{a}
    \right)^k, \quad k \approx 1, \label{2.3}
\end{equation}
where $L_{cr}$ is the critical luminosity above which a RS arises (see, e.g., \citealt{Mushtukov2015a}) and $a$ is the characteristic shock height at about the critical luminosity. We note that this height corresponds roughly to the transverse size of the polar cap. For concreteness, we take the radius of the circular shock to be $a_0 = a$.
As \cite{Musthukov2015b} argued, the effective deceleration of the accreting plasma for sub-critical luminosities occurs at an altitude comparable to the radius of the polar cap. Thus, it is sound to assume that an appropriate length scale for the shock height at the critical luminosity is the size of the polar cap $a$.

Combining Eq. \eqref{2.3} with Eq. \eqref{2.2}, one can obtain an analytical formula for the expected $E_c - L_X$ anti-correlation. \cite{Poutanen2013} remarked that this expectation is inconsistent with the observed correlation $E_{\rm CRSF} - L_X$ in the source V0332+53, mainly because the theoretical rate of change of $E_c$ with $L_X$ is too large. However, this thinking does not account for the effects discussed below.

\subsection{Relativistic corrections}
\label{sec2.2}

\subsubsection{Doppler effect}
\label{sec2.2.1}

In \cite{Loudas2023}, we showed that the Doppler effect plays a crucial role in the line-forming process, mainly by red-shifting the line centroid. This effect is reduced for larger heights $H$, because the free-fall velocity (bulk-motion speed)  decreases with $H$. Thus, the Doppler effect is expected to affect the dependence of the CRSF centroid on luminosity.

Neglecting all quantum mechanical corrections, the resonant energy (i.e., where the peak of the scattering cross-section is) equals the classical cyclotron energy $E_c$. Thus, photons of energy $E_c$, as measured in the accreting plasma's frame (bulk-motion frame), get resonantly scattered. However, their energy is different at the RS frame (the lab frame), due to the mildly relativistic velocity of the bulk-motion frame. As a result, photons that undergo resonant scattering with the in-falling electrons have lower energies (as measured in the lab frame) with respect to $E_c$, thereby the observed CRSF does not appear at $E_c$ in the emergent spectrum, but at a lower energy. Below, we quantify this effect.

Assume a photon of energy $E$, as measured in the RS reference frame (i.e., the lab frame), moving at an angle $0 < \theta < \pi$ with respect to the magnetic field axis (although photons are initially injected from the shock toward the pre-shock region, implying $\theta\in[0,\pi/2)$, the transverse Thomson optical depth is larger than one and therefore most photons are scattered multiple times losing information regarding the angular distribution of the
initial emission -- thus we consider all possible values of $\theta$), and scattered by a free-falling electron in the pre-shock region, which is moving with a velocity
\begin{equation}
    v_{ff}(H) = - c\sqrt{\dfrac{r_s}{R_* + H}}. 
    \label{2.4}
\end{equation}
where $r_s=2GM_*/c^2$ is the Schwarzschild radius, $G$ is the gravitational constant, and $M_*$ is the mass of the NS.
Applying a Lorentz transformation from the RS frame to the bulk-motion frame, the incident photon's energy in the bulk-motion frame equals
\begin{equation}
    E^\prime = \gamma_{ff} E (1 + \beta_{ff} \cos\theta), 
    \label{2.5}
\end{equation}
where $\beta_{ff} = |v_{ff}| / c $ and $\gamma_{ff} = 1 \big/ \sqrt{1 - \beta_{ff}^2}$.

Demanding the photon to undergo resonant scattering, we set $E^\prime = E_c$.
Moreover, we can average the right hand side of Eq. \eqref{2.5} over all solid angles using an appropriate distribution function $f(\theta)$ to obtain the line centroid energy. Given that the angular part of the magnetic resonant cross-section (see \citealt{Loudas2021}) is, to leading order, proportional to $1 + \cos^2\theta$, we choose to employ the following distribution function
\begin{equation}
    f(\theta) = \dfrac{3}{8} (1 + \cos^2\theta),
\end{equation}
where $\int_{-1}^1 f(\theta) ~ \der\cos\theta = 1$. So, Eq. \eqref{2.5} becomes 
\begin{equation}
    E_c = \tilde{E}_c \left<\gamma_{ff} (1 + \beta_{ff} \cos\theta)\right>_f, 
\end{equation}
where $\tilde{E}_c$ refers to the centroid of the CRSF that is formed at the RS, as measured in its frame (i.e., in the lab frame), and
\begin{equation}
    \left<\gamma_{ff} (1 + \beta_{ff} \cos\theta)\right>_f = \gamma_{ff} \int_{-1}^{1} (1 + \beta_{ff} \cos\theta) f(\theta) ~ \der\cos\theta. 
\end{equation}
It is trivial to show that 
\begin{equation}
\left<\gamma_{ff} (1 + \beta_{ff} \cos\theta)\right>_f = \gamma_{ff}.
\end{equation}
Eventually, we end up with
\begin{equation}
    \tilde{E}_c = \dfrac{1}{\gamma_{ff}} E_c.
    \label{2.6}
\end{equation}

Therefore, the correction term due to the Doppler shifting is
\begin{equation}
    {\cal D} = \dfrac{1}{\gamma_{ff}}.
    \label{2.7}
\end{equation}
This correction has two implications: 1) the line centroid always appears at lower energies compared to the classical cyclotron energy $E_c$, in line with results of MC simulations \citep{Loudas2023}; 2) the change of the line-energy centroid with luminosity (RS height) does not only arise from the magnetic-field variation along the column, but also has an additional dependence through the free-fall velocity ($\gamma_{ff}$ factor; see Eq. \ref{2.4}).  

\subsubsection{Gravitational redshift}
\label{sec2.2.2}

Another relativistic correction, that adds an extra dependence of $E_{\rm CRSF}$ on luminosity $L_X$, that should be taken into account when fitting data, is the gravitational redshift. In Sect. \ref{sec2.2.1}, we showed that the cyclotron line in the X-ray spectrum that emerges from the RS is centered at $\tilde{E}_c$ (see Eq. \ref{2.6}). However, the observed line centroid $E_{\rm CRSF}$ will appear to a distant observer at lower energies owing to the gravitational redshift. Considering a Schwarzschild metric outside the NS, the observed cyclotron-line energy is   

\begin{equation}
    E_{\rm CRSF} = \dfrac{1}{1 + z} \tilde{E}_c, 
    \label{2.8}
\end{equation}
where $z$ is the gravitational redshift
\begin{equation}
    z = \sqrt{\dfrac{1}{1 - \dfrac{r_s}{R_{*} + H}}} - 1 = \gamma_{ff} - 1.
    \label{2.9}
\end{equation}
In the right part of the above equation, we used Eq. \eqref{2.4} along with the definition of $\gamma_{ff}$.

Hence, we find that the gravitational correction is 
\begin{equation}
    {\cal G} = \dfrac{1}{\gamma_{ff}}.
    \label{2.10}
\end{equation}

\subsection{Predicted dependence of $E_{\rm CRSF}$ on $L_X$}

Starting from Eq. \eqref{2.2} and incorporating the relativistic corrections ${\cal D},~{\cal G}$, we get the following revised formula for the dependence of the {\it observed} cyclotron line centroid on the height $H$ of the RS
\begin{align}
    E_{\rm CRSF}(H)
    = 
    {\cal G} \times {\cal D} \times E_c(H) 
    &= 
    \dfrac{1}{\gamma_{ff}^2} \left(\dfrac{1}{1 + H/R_*}\right)^3 E_{c,*} \notag \\
    &= 
    \left(1 - \dfrac{r_s}{R_* + H}\right) \left(\dfrac{1}{1 + H/R_*}\right)^3 E_{c,*} .
    \label{2.11}
\end{align}
Expanding this formula for small $H/R_*$ and keeping up to first order in $H/R_*$, we find
\begin{equation}
    \dfrac{\Delta E_{\rm CRSF}(H)}{E_{c,*}} \equiv \dfrac{E_{\rm CRSF}(H)-E_{c,*}} {E_{c,*}} \approx - \dfrac{r_s}{R_*} - \left(3 -4\dfrac{r_s}{R_*}\right)\dfrac{H}{R_*} .
    \label{2.12}
\end{equation}
The terms proportional to $r_s/R_*$ correspond to the relativistic corrections. Clearly, the rate of change of $E_{\rm CRSF}$ with $H/R_*$ is $3 - 4(r_s/R_*)$, where the term 3 is due to the rate of change of the strength of the magnetic dipole with $H/R_*$. For a typical NS, $r_s/R_* \sim (0.4-0.5)$, thus the predicted rate of change for small $H/R_*$ becomes $\sim (1-1.4)$, which is significantly smaller than $3$. The relativistic corrections {\it decrease} the naive rate of change expected from the magnetic-dipole variation with height $H$. 

Using Eq. \eqref{2.3} to connect the height $H$ of the RS with the X-ray luminosity and substituting it into Eq. \eqref{2.11}, one can obtain the theoretically predicted $E_{\rm CRSF} - L_X$ anti-correlation. The free parameters in this model are five: $E_{c,*}, ~ L_{cr}, ~ a/R_*, ~ r_s/R_*, \text{and} ~ k$.

\section{Results} 
\label{sec3}

\begin{figure}
        \centering \includegraphics[width=9cm]{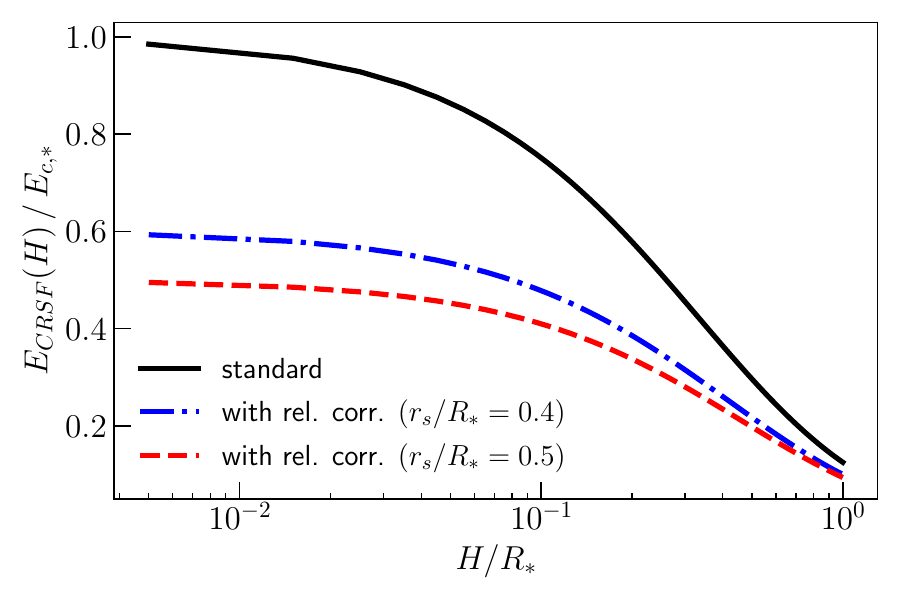}
        \caption{Predicted variation of the $E_{\rm CRSF}(H)$ value (normalized to $E_{c,*}$) with the shock front's altitude $H$ (normalized to $R_*$). The black solid line refers to the standard picture $r_s=0$ (no relativistic corrections). The blue dot-dashed line includes relativistic corrections with $r_s/R_*=0.4$, while the red dashed line stands for $r_s/R_* = 0.5.$}
        \label{fig1}
    \end{figure}

To demonstrate the effect of the relativistic corrections on the predicted variation of the CRSF energy centroid along the column, we plot in Fig. \ref{fig1} the theoretically predicted $E_{\rm CRSF}(H)$ (normalized to $E_{c,*}$) against the shock front's altitude $H$ (normalized to the NS radius $R_*$). To calculate $E_{\rm CRSF}$ as a function of $H$, we employ Eq. \eqref{2.11}. The black solid line corresponds to the standard prediction of the line energy variation due to the magnetic-field strength without relativistic corrections, that is, with $r_s = 0$. The red (blue) dashed (dot-dashed) line refers to the theoretically predicted cyclotron-line variation including the relativistic corrections discussed above, with $r_s/R_* = 0.5$ ($r_s/R_* = 0.4$). For a typical NS, $r_s/R_* = 0.42$.

As it can be inferred from Fig. \ref{fig1}, the effects of relativistic corrections are two: 1) The rate of change of the line energy with the shock's height is significantly lower than the one due to only the magnetic-field strength variation. The relativistic terms reduce the dependence of the cyclotron line energy on the shock's height.  2) {\it The observed CRSF energy is significantly smaller than the cyclotron-line energy $E_{c,*}$ on the surface of the NS}, depending on the value $r_s/R_*$ (see the first term in Eq. \ref{2.12}).  The larger the compactness ($r_s/R_*$) of the NS is, the greater the deviation of the observed CRSF from $E_{c,*}$ will be for a given shock's height. Therefore, {\it relativistic corrections must be taken into account, if one demands a robust estimation of the magnetic field strength on the NS's surface from the imprinted CRSF on the emergent X-ray spectrum.}

In view of the above, it is inviting to try to fit the {\it observed} anti-correlation for the X-ray source V0332+53 \citep{Tsygankov2010}, a source residing in the super-critical luminosity
regime. Since the data are of the form $(L_X, ~ E_{\rm CRSF})$, our model is the combination of Eq. \eqref{2.11} with Eq. \eqref{2.3}. Based on a previous analysis by \cite{Becker2012}, the critical luminosity $L_{cr}$ for this source is approximately $L_{cr}\approx 4\times 10^{37}$ erg s$^{-1}$. We fix this parameter to be $L_{cr} = 4\times 10^{37}$ erg s$^{-1}$.
We also fix the power-law index $k$ (shown in Eq. \ref{2.3}) to be $k=1$ (see \citealt{Basko1976}). Therefore, we end up with only three free parameters. We note that we only fit data with luminosity higher than the chosen critical one, as the model is valid only in this regime.

\begin{figure}
        \centering \includegraphics[width=9cm]{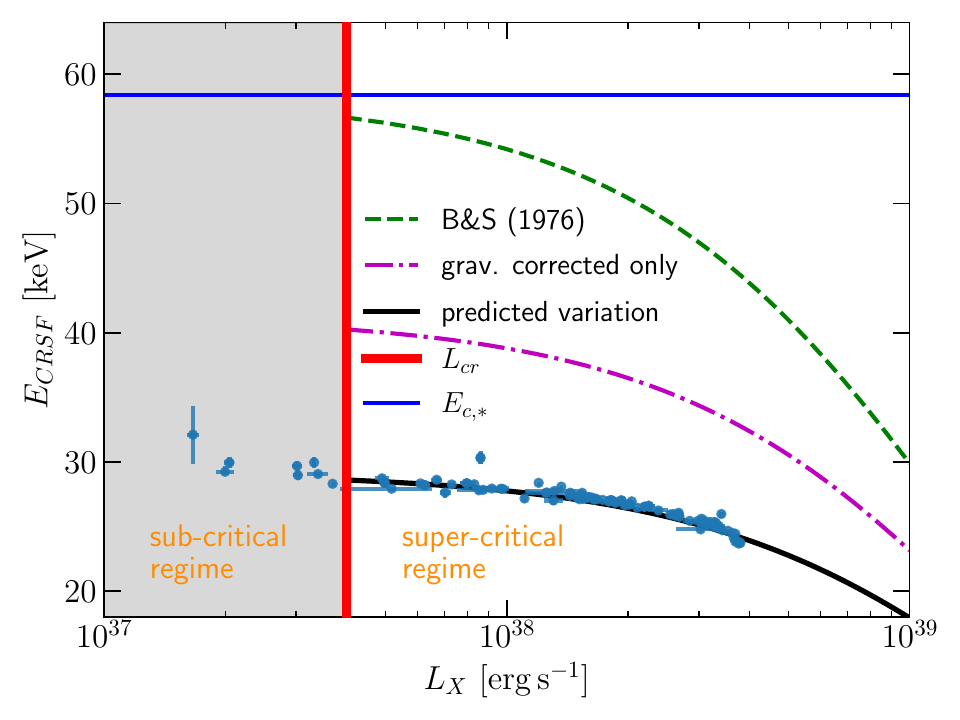}
    \caption{Observed cyclotron-line energy variation of the source V0332+53 with luminosity. The blue points correspond to data reported in \cite{Tsygankov2010}. The black solid line is the optimal fit to the data, using our theoretical prediction. The green dashed line is the expected variation only due to the magnetic-field strength (see Eq. \ref{2.2} and \citealt{Basko1976}), using as $E_{c,*}$ and $a$ the parameters obtained through the fit. The magenta dash-dotted line is the one accounting for the magnetic field variation and gravitational redshift, but not the Doppler effect (see Sect. \ref{sec2.2.1}). The red vertical line denotes the critical luminosity, and the blue horizontal line represents the cyclotron line energy on the NS's surface.}
        \label{fig2}
    \end{figure}

    The results are displayed in Fig. \ref{fig2}, while the optimal parameters are given in Table \ref{tab1}.
    The blue points correspond to the data obtained by \cite{Tsygankov2010}, the red vertical line indicates the chosen critical luminosity, and the blue horizontal line denotes the inferred (i.e., the actual) cyclotron-line energy on the NS's surface.
The black solid line displays the model's best fit to the data, with parameters that are shown in Table \ref{tab1}, and the green dashed line is the expected variation due to the magnetic-field strength without any relativistic correction (\citealt{Basko1976}; see Eq. \ref{2.2}), where we have used the values of the parameters $E_{c,*}$ and $a/R_*$ that are given in Table \ref{tab1}, while the magenta dash-dotted line accounts for the magnetic field variation and gravitational redshift only. An eye examination of our model fit to the data (solid line in Fig. \ref{fig2}) shows that the model fits the data impressively well. We remark that the strength of the magnetic field on the surface of the neutron star is about double of what one would infer from a naive look at the observed CRSF at the critical luminosity.

    \begin{table}
        \caption{Optimal parameters.}
        \label{tab1}
        $$
        \begin{array}{p{0.4\linewidth}l}
            \hline
            \noalign{\smallskip}
            parameter  &      {\mathrm{value}}  \\
            \noalign{\smallskip}
            \hline
            \noalign{\smallskip}
            $E_{c,*} \,\,{\text{[keV]}}$  & 58 \pm 8    \\
            $r_s/R_*$      &  0.49 \pm 0.07 \\
            $a/R_*$     & 0.011 \pm 0.001             \\
            \noalign{\smallskip}
            \hline
        \end{array}
        $$
   \end{table}

    \section{Discussion and conclusions} 
    \label{sec4}
	
	In this study, we investigated the predicted variation of the CRSF energy centroid, that is formed at the radiative shock in the accretion column of an accreting magnetic NS, with the shock's height $H$ from the NS surface. In \cite{Loudas2023}, we examined the proposed model of \cite{Basko1976} and demonstrated that prominent cyclotron line features can form in the radiative shock of an accreting NS through magnetic resonant scattering. We further highlighted that the Doppler effect between the accretion-flow's frame (bulk-motion frame) and the shock's frame (lab), as well as the effect of the gravitational redshift, are important for the determination of the energy $E_{\rm CRSF}$, at which the CRSF appears in the spectrum.  Both effects shift the CRSF energy centroid to values {\it lower} than the local cyclotron energy $E_c(H)$.
 
    We quantified the effects of both the Doppler boosting and the gravitational redshift in the observed cyclotron-line centroid and obtained an analytical formula that describes the variation of the $E_{\rm CRSF}$ with the X-ray luminosity $L_X$. {\it We found that relativistic corrections noticeably decrease the expected rate of change of the CRSF energy centroid with the shock's height, therefore alleviating the discrepancy between the naively expected (i.e., without relativistic corrections) $E_c - L_X$ anti-correlation and the observed one $E_{\rm CRSF} - L_X$ in XRPs of supercritical luminosity.}

    We remark that \cite{Becker2012} claim that they were able to explain the cyclotron-line variation with luminosity in V0332+53.  Since they did not consider any of the relativistic effects described above, it seems that they were able to do so because they considered that the cyclotron line is formed {{\it well below the radiative shock}}, in the post-shock region, and relatively close to the surface of the neutron star. As a matter of fact, for pulse-to-pulse variations, they inferred that at about the critical luminosity, the height of the region emitting the cyclotron line is $\sim 12$ m (see their Eq. 40 and Table 2). This means that, at ten times the critical luminosity, the emitting region is at $\sim 120$ m, and the dipole field there has not changed much. 
    This, however, is not justified for three reasons. 
    1) Because the radiative shock is thin \citep{Loudas2023}, 
    most of the accretional energy ($\sim 98\%$; assuming a velocity decrease across the shock of a factor of $\sim 7$, appropriate for radiative shocks) is emitted at the shock, and very little accretional energy ($\sim 2\%$) is emitted at the post-shock region. This is why we consider in this study that the CRSF is formed at the shock altitude. In addition, there is no characteristic height in the post-shock region, which one could consider as the site of preferential formation of the cyclotron line. 
    2) It has been shown by \cite{Mushtukov2015a} that, at the critical luminosity, the accretion mound has a height comparable to the size of the footprint of the accretion column.  Thus, at the critical luminosity, the shock has a height at least equal to the height of the accretion mound.
    3) Recent Radiation Magneto-Hydrodynamic (RMHD) simulations \citep{Zhang2022} support the idea that the cyclotron line is emitted at the radiative shock, because most of the radiation escapes at the shock and not far below it (see also \citealt{Zhang2023, Sheng2023}). Although the column below the shock is hot, the total emergent flux is dominated by sideways emission at the shock altitude, owing to the fact that the accretional energy is liberated at the shock (see, e.g., Fig. 6 \& 16 in \citealt{Zhang2022}).
    
    \cite{Poutanen2013} argued against the cyclotron line formation in the RS model based on the tension between the weak observed and the theoretically steep $E_{c}-L_X$ anti-correlation, though ignoring the relativistic corrections. We demonstrated here that relativistic effects resolve this puzzle. Moreover, the reflection model \citep{Poutanen2013} predicts a convex function describing the $E_{\mathrm{CRSF}}-L_X$ negative correlation and thus fails to fit the data over the total luminosity range (see Fig. 5 in \citealt{Poutanen2013}). In contrast, the shock model predicts a concave function, that is the second derivative of $E_{\mathrm{CRSF}}$ with respect to $L_X$ is negative, and fits the data over the full range of luminosity once relativistic corrections are included. All observed anti-correlations (see, e.g.,  Fig. 3 in \citealt{Becker2012}) seem to be described by concave functions, in line with the shock model.
    
    Another factor that could be important in this analysis is the relation between the X-ray luminosity and the shock front's height (see Eq. \ref{2.3}). The luminosity scales approximately linearly with height (i.e., $k=1$) as long as the cross sectional area of the column does not vary noticeably. One could potentially argue, though, that the radiation pressure exerted on the sidewall of the accretion column, especially taking into account the resonant cross section, can be high enough at the high-luminosity regime, to push plasma sideways and expand the transverse size of the column. This could effectively increase the index $k$ in Eq. \eqref{2.3} and weaken even more the $E_{CRSF} - L_X$ predicted anti-correlation, apparently affecting the values of fitted parameters shown in Table \ref{tab1}.
    However, as we demonstrate in Appendix \ref{App.A}, the magnetic pressure is the dominant one even for super-critical luminosities and therefore this effect can be safely ignored in our considerations.
    Nevertheless, RMHD simulations \citep{Zhang2022} in a magnetic column geometry at the very high-luminosity regime could address this effect in detail and provide robust conclusions.

    Here, we used the 1D solution of \cite{Basko1976}, which suggests a stationary, plane-parallel radiative shock the height of which scales linearly with luminosity (i.e., $k=1$ in Eq. \ref{2.3}). \cite{Zhang2022} performed robust 2D RMHD simulations and found that the shock exhibits high-frequency oscillations, its shape is parabolic, and the time-average height of the shock front is lower than the one predicted from \cite{Basko1976} for given accretion rate (luminosity), due to the larger cooling efficiency of the column in 2D. To which extent the oscillations of the shock affect our conclusions is difficult to determine without studying the spectral formation in a time-dependent magnetic column dynamical structure. Furthermore, one could relax the condition $k=1$ in our consideration and explore larger values, i.e., weaker luminosity-height relations as the findings of \cite{Zhang2022} imply. This is though outside of the scope of this paper, which primarily aims to highlight the relativistic effects on the predicted cyclotron line variation with luminosity. Finally, even if the shock has a paraboloidal-like profile, our plane-parallel shock approximation still holds for our purposes, provided most of the radiation escapes sideways at an altitude roughly equal to the height of the shock front in the middle of the column (see Fig. 16 in \citealt{Zhang2022}) and thus it is the appropriate height to consider as the altitude of the line-forming region for an analytical model like the one we offer in this study.
    
    In summary, we showed that, contrary to recent common belief, simple Physics can explain analytically the anti-correlation between $E_{\rm CRSF}$ and $L_X$ for the source  V0332+53.

	\begin{acknowledgements}
        We would like to thank the anonymous referee for providing a constructive report that helped us improve the quality and clarity of this manuscript.
        We are indebted to R\"udiger Staubert for sending us the observational data of the X-ray source V0332+53, originally reported in \cite{Tsygankov2010}. 
        NL would like to thank Lizhong Zhang for fruitful and stimulating discussions related to the accretion column physics.
        NL acknowledges support from the European Research  Council (ERC) under the HORIZON ERC Grants 2021 Program, under grant agreement No. 101040021 and funding from Princeton University First-Year Fellowship in the Natural Sciences and Engineering.
	\end{acknowledgements}

	\bibliography{references}
  
\begin{appendix}

\section{Radiation pressure on the sidewall}
\label{App.A}

In this appendix, we study the radiation pressure exerted on the outer layer of the magnetic column due to X-rays emitted at the shock. The magnetic field has a leading role in preserving the shape of the accretion column (axisymmetric in our case), yet radiation created in the shock region results in extraordinary values of radiation pressure on the wall, provided the shock is thin. 

It is assumed that for an accreting NS of super-critical luminosity, the height of the radiative shock depends on the mass-accretion rate and scales approximately linearly with the emergent X-ray luminosity (e.g., \citealt{Lyubarskii1988}). However, this relation is based on the assumption that the transverse size of the accretion column does not vary with luminosity. This is correct, if one accounts only for the radiation pressure due to photons emitted below the shock. However, radiation that emerges from the shock region carries away the majority of the accretional energy and it escapes almost perpendicular to the column, thereby exerting a much higher pressure on the sidewall than the photons below the shock. Suppose this pressure is comparable to or greater than the magnetic pressure. In that case, the shock's radiation can expand the transverse size of the accretion column, instead of only raising the shock, thus lowering the rate of the shock's height increase with luminosity. Such a scenario could, in principle, alleviate the tension between the observed $E_{\rm CRSF} - L_{X}$ negative correlations and the theoretically expected ones (ignoring relativistic corrections). Whether this argument holds is explored below, where we  compute the ratio of the radiation pressure on the column's wall to the magnetic pressure, to check if the flux tube in which accretion takes place has a fixed shape.

\subsection{Radiation pressure}

We demonstrated in \cite{Loudas2023} that, for high-luminosity sources, the shock's width $\zeta$ is significantly smaller than its size $a$ across the column. So, it is plausible to work under the thin-disk approximation to describe the shock. We consider for simplicity that all the emergent radiation is emitted from the shock region and escapes sideways, since the optical depth along the column is effectively infinite (see \citealt{Loudas2023}). We treat the shock as a radiating disk of radius $a$ that emits photons in the plane of the disk. This assumption overestimates the amount of radiation that escapes sideways.  Thus, our calculation of the radiation pressure on the wall will yield an upper limit on the efficiency of radiation to push outward the boundary of the accreting flux tube.

Under the aforementioned assumptions, one can relate the total emerging luminosity $L$ to the total flux $F_\perp$ perpendicular to the column through the following equation
\begin{equation}
	F_\perp = \dfrac{L}{{S}} = \dfrac{L}{2\pi a \zeta}, \label{8.1}
\end{equation}  
where  $S$ is the cylindrical area of the disk.  Exploiting the axial symmetry of the problem, we may compute only the radiation pressure on the wall along an axis, say the $x$-axis. Under the radiation diffusion approximation, the radiation pressure gradient reads (see, e.g., Eq. 4 of \citealt{Mushtukov2015c}) 
\begin{equation}
	\dfrac{\der p_R(x)}{\der x} = \dfrac{n_e}{c} \int_0^\infty \sigma_\perp(E) F_{E,\perp}(x) ~\der E,
	\label{8.2}
\end{equation}
where $p_R(x)$ is the radiation pressure at $x$, $n_e$ is the electron number density (assumed constant) in the shock, $\sigma_\perp$ is the scattering cross-section for the scattering of a photon propagating perpendicular to the accretion column with an electron, and $F_{E,\perp}(x)$ is the differential radiation flux in the shock as a function of energy and radial position $x$. Following \cite{Mushtukov2015c}, we consider for simplicity that $F_{E,\perp}$ increases linearly across the shock, that is 
\begin{equation}
    F_{E,\perp}(x) \approx \dfrac{x}{a}F_{E,\perp}(a), \label{8.2new}
\end{equation}
where $F_{E,\perp}(a)$ refers to the differential radiation flux on the sidewall and therefore the total radiation flux $F_\perp$ (escaping the column perpendicularly) is 
\begin{equation}
F_{\perp} = \int_0^\infty F_{E,\perp}(a) ~\der E. \label{8.3new}
\end{equation}
Substituting \eqref{8.2new} into \eqref{8.2} and integrating over x yields
\begin{equation}
	P_{R} \approx \dfrac{n_e a}{2c} \int_0^\infty \sigma_\perp(E) F_{E,\perp}(a) ~\der E, \label{8.3}
\end{equation}
where $P_R \equiv p_R(a)$ is the radiation pressure on the wall, while we used $p_R(x=0) = 0$.

Following the work of \cite{Mushtukov2015a} (see Eq. 2 therein), the right-hand side (RHS) of \eqref{8.3} can be treated as follows: we integrate over energy assuming the cross-section to be independent of the energy (though it is not the case) and then substitute it with an effective cross-section $\sigma_{eff}$ (for the definition of $\sigma_{eff}$ see Appendix \ref{App.A3}), namely 

\begin{equation}
	P_R \approx \dfrac{n_e \sigma_{eff}a}{2c} \int_0^{\infty} F_{E,\perp}(a) ~\mathrm{d}E  = \dfrac{\tau_\perp}{2} \dfrac{F_\perp}{c} \left(\dfrac{\sigma_{eff}}{\sigma_\tau}\right), \label{8.4}
\end{equation}
where we used $\tau_\perp = n_e \sigma_T a$ and Eq. \eqref{8.3new}. Finally, taking into account \eqref{8.1}, we end up with
\begin{equation}
	P_{R} \approx \tau_\perp \left(\dfrac{\sigma_{eff}}{\sigma_\tau}\right)\dfrac{L}{4\pi a \zeta c}.\label{8.5}
\end{equation}

\subsection{Comparison to the magnetic pressure} \label{secB.2}

The magnetic pressure is equal to
\begin{equation}
	P_B = \dfrac{B^2}{8\pi}, \label{8.6}
\end{equation}
where $B$ is the magnetic field strength at the shock's height. The ratio of the radiation pressure on the wall to the magnetic pressure reads
\begin{equation}
	\dfrac{P_R}{P_B} \sim 2 \tau_\perp  \left(\dfrac{\sigma_{eff}}{\sigma_\tau}\right)\dfrac{L}{a \zeta c} B^{-2},\label{8.7}
\end{equation}
where we employed Eq. \eqref{8.5}. To find out whether the radiation produced at the shock is able to push the material of the accretion column into field lines outside it, and therefore to affect the shock-height - luminosity relation, derived for a fixed accretion-column size, we make use of the approximate scaling relations obtained in Sect. 2 of \cite{Loudas2023} and deduce
\begin{equation}
	\dfrac{P_R}{P_B} \sim 1.2 \times 10^{-6}
	\left(\dfrac{\sigma_{eff}}{\sigma_\tau} \right)
	\left(\dfrac{\dot M}{10^{16}~\mathrm{gr ~ s^{-1}}} \right)^{11/5}	\left(\dfrac{B}{10^{12}\, \mathrm{G}}\right)^{-1}.\label{8.8}
\end{equation}  

In the next section, we set an upper bound on $\sigma_{eff}$ and therefore on the ratio $P_R/P_B$.

\subsection{An application to a typical accreting NS} \label{App.A3}
For a typical accreting NS of super-critical accretion luminosity, say $L \approx 10^{38} ~\mathrm{erg ~s^{-1}}$, that is a mass-accretion rate $\dot M \approx 5\times 10^{17} ~\mathrm{g~s^{-1}}$, and a magnetic-field strength on the surface $B \approx 5 \times 10^{12}$ G, the ratio of the radiation pressure to the magnetic one, according to \eqref{8.8}, is 
\begin{equation}
	\dfrac{P_R}{P_B} \sim 1.3 \times 10^{-3} \left(\dfrac{\sigma_{eff}}{\sigma_\tau} \right). 
    \label{8.9}
\end{equation}
To estimate the effective cross-section,
we should allow for resonant scattering, which significantly enhances the probability of interaction between matter and radiation for energies close to the cyclotron energy (see, e.g., \citealt{Loudas2021}). The cross-section for energies around the resonant frequency and magnetic fields on the order of a few $10^{12}\,\mathrm{G}$ can reach up to values $\sim {\cal O}(10^5\sigma_\tau)$. But, the energy range around the cyclotron energy that the resonant scattering is efficient is roughly equal to the classical cyclotron line-width $\Gamma$, which is less than $0.5$ keV for typical NS conditions. So, assuming for simplicity that the X-ray radiation spectrum, which emerges from the shock, lies in the range $(1-100)$ keV, we can revisit Eq. \eqref{8.4} and estimate the effective cross-section $\sigma_{eff}$ 
using the following formula
\begin{equation}
\sigma_{eff} \approx \sigma_\tau\dfrac{\int \sigma_{res,\perp}(E) F_{E,\perp}(a)~\der E}{\int \sigma_\tau F_{E,\perp}(a) ~\der E},
\label{8.11}
\end{equation}
where $\sigma_{res,\perp}$ is the resonant cross-section for $e-\gamma$ interaction of a photon moving perpendicular to the magnetic field, while the integral is over the energy range $(1-100)$ keV. Suppose the differential radiative flux has a power-law like form
\begin{equation}
    F_{E,\perp} \propto E^{-\kappa}, \label{8.12}
\end{equation}
we then perform the integral numerically and obtain
\begin{equation}
	\sigma_{eff}\approx 2.1 \times 10^1 ~ \sigma_\tau, \quad \text{for} ~~ \kappa=2,\label{8.13}
\end{equation} 
and
\begin{equation}
	\sigma_{eff}\approx 0.3 ~ \sigma_\tau, \quad \text{for} ~~ \kappa=4.\label{8.14}
\end{equation} 
Substituting these values into \eqref{8.9}, we find 
\begin{equation}
    \dfrac{P_R}{P_B} \lesssim ~ 2.7 \times 10^{-2}, \quad \text{for} ~~ \kappa=2,
    \label{B.11}
\end{equation}
and
\begin{equation}
    \dfrac{P_R}{P_B} \lesssim ~ 3.9 \times 10^{-4}, \quad \text{for} ~~ \kappa=4.
    \label{B.11}
\end{equation}
We conclude that a hard X-ray energy spectrum, e.g., $\kappa=2$, yields a radiative pressure that is way smaller than the magnetic pressure, even in the extreme scenario where all radiation escapes perpendicular to the shock. For softer spectra, e.g., $\kappa=4$, the pressure ratio decreases significantly since fewer photons have energies that lie in the narrow resonant scattering energy range. 

We remark that, even if resonant scattering is due to thermal electrons and not bulk-motion ones, the inferred pressure ratio does not change significantly. Using the thermally averaged cross-section (see Appendix of \citealt{Loudas2023}), we found that the effective cross-section does not vary by more than $20\%$ (even for $kT_e=50$ keV), implying that thermal effects are not important in our consideration. The main reasons are two: 1) the thermal broadening of the resonant cross-section is minimum for photons moving perpendicular to the motion of the electrons (see Appendix of \citealt{Loudas2023}), as the Doppler effect is not that important in this case, and 2) although the thermal broadening yields a greater cyclotron-line width $\Gamma$, and therefore a larger fraction of photons can be resonantly scattered by electrons, the peak of the cross-section decreases, making the resonant optical depth smaller. 

Therefore, we safely place this upper limit on the ratio of the radiation pressure to the magnetic pressure 
for conditions similar to the ones regularly observed in accreting NSs and rule out the possibility of radiation pressure expanding the column size.  

\end{appendix}

\end{document}